\documentclass[12pt]{iopart}
\usepackage{iopams}  
\usepackage{color}
\usepackage{tabularx}
\usepackage{epsfig}
\usepackage{graphicx}
\begin{document}
\vspace {1.0 in}

\title{A Simple Spin Glass Perspective on Martensitic Shape-memory Alloys}

\footnote{This paper is dedicated to John Pendry in celebration of his 65th birthday.}

\author{David Sherrington}
\address{Rudolf Peierls Centre for Theoretical Physics, University of Oxford,
1 Keble Road, Oxford OX1 3NP, UK \footnote{Permanent address},
\\
Center for Nonlinear Studies, Los Alamos National Laboratory, Los Alamos, NM 87545, USA
\\
{\rm and}
\\
Santa Fe Institute, 1399 Hyde Park Rd., Santa Fe, NM 87501, USA}

\ead{D.Sherrington1@physics.ox.ac.uk}


\date{3 January 2007}

\begin{abstract}
A brief qualitative mapping is given between austenite, 
tweed and twinned phases of martensite alloys and corresponding 
paramagnetic, spin glass and periodic phases in spin glass alloys.  
\end{abstract}
\maketitle

\section{Introduction}
Ferroelastic martensitic alloys show discontinuous phase 
transitions as the temperature is 
lowered from a high temperature high-symmetry phase to 
a low temperature 
phase of twinned lower-symmetry variants; for example
from an austenite phase with cubic 
symmetry to an ordered pattern of stripes of
complementarily distorted tetragonal twins \cite{Bhattacharya}.
Often these two phases are separated by an intermediate phase, known as 
"tweed" in which the tetragonal twins are arranged in a more random way \cite{Tanner}. 
The lower-temperature phases exhibit interesting shape-memory and aging behaviour \cite{Bhattacharya, Otsuka}.

In this paper we present a simple
caricaturization of these phases and their transitions and
features, in terms of simple 
spin glass alloy models. The suggestion that tweed is an analogue of
a spin glass and a description of these systems 
in terms of a spin glass model is not new \cite{Kartha91,Sethna92,Kartha92},  
but the modelling here is somewhat different and, hopefully, instructive. 
It also suggests interesting new models (or variations) for study
by spin glass physicists.

\section{Modelling}

The origin of the structural transitions above  
is viewed as due to the frustrated interplay of competitive 
interactions of different ranges and anisotropy 
\cite{Lookman03}, combined 
with quenched disorder due to statistical 
inhomogeneity in the alloy make-up \cite{Kartha91}. The origin of the competing 
forces is elastic but, rather than retaining the 
technical complications of a complete elastic analysis,
we shall here pursue a much 
simpler mean-field caricaturization in terms of effective `spins', 
with the aim of conceptual 
simplification and qualitative explanation and prediction.

Our starting point is to separate the local and effective interactive 
features of the martensitic transition. The transitions 
from austenite to tetragonal are observed as first order and 
thus we may view the local propensity to either cubic or 
tetragonal symmetry as describable by a first-order
Landau transition. 

\subsection{Two dimensions}

In a three-dimensional cubic system there are three mutually orthogonal tetragonal 
local distorsion orientations. However, for simplicity we shall initially consider a 
two-dimensional system in which  
there are  only two possible rectangular variants, orientated 
with their longer axes 
in the $x$ or the $y$ direction. They may be 
characterised by a variable
$\phi$ whose magnitude indicates the strength of tetragonal distorsion and
whose sign indicates which of the two orientations it is in; explicitly $\phi$ 
is the deviatoric strain $\phi = (\epsilon_{11} -\epsilon_{22})/\sqrt{2}$ where 
$\epsilon_{ij}$ is a component of the Lagrangian strain tensor \cite{Lookman01}. 
The local situation can thus be described 
in terms of a local free-energy function
\begin{equation}
F_0=\sum_{i} [{ A_{i}(T) \phi_{i}^2 - B_{i}(T)\phi_{i}^4 + C_{i}(T)\phi_{i}^6}]
\label{eq:local_free_energy}
\end{equation}
where  the $B$ and $C$ are both positive and their temperature 
dependence is not of qualitative consequence, while the $A(T)$ 
change sign from positive to negative as the temperature $T$ is 
reduced.  $i$ labels the local region, and the $A, B$, and  $C$ are 
given subscipts $i$ to allow for quenched heterogeneity.  
Minimizing the free energy leads to a transition 
from $\phi_i = 0$ to $\phi_{i} = \pm \phi{_{i}}^*$ as $T$ is 
reduced. 
$\phi=0$ is interpreted as the locally cubic structure 
and the $\pm \phi_{i}^*$ as the two tetragonal distorsions.
The interesting behaviour arises from the inhomeneity in the 
passage of the $A_i(T)$ across zero. The signs of $B$ and $C$ are 
chosen to reproduce the observed first-order transitions. 
The
$i$- and $T$-dependences of the $\phi_{i}^*$ are
not of qualitative importance and so will be ignored henceforth.
  
In a further simplification we may re-write $F_0$ as an effective 
`spin Hamiltonian' \cite{Lookman04}.
\begin{equation}
H_0=\sum_{i} D_i(T) S_{i}^2; \qquad S_{i} = 0,\pm 1
\label{eq:GS_local_free_energy}
\end{equation}
with $S=0$ corresponding to a local cubic structure, $S=\pm 1$ 
corresponding to the two orthogonal rectangular distorsions and
the $D_i$ changing from positive to negative as the 
temperature $T$ is reduced, emulating the sign-change of $A_i(T)$.
The ground state has $S_i=0$ (cubic) for $D_i\geq 0$ and $S_i=\pm 1$ 
(tetragonal) for $D_i\leq 0$.
Below we shall continue to investigate the effect of the variation 
of the $D_i$ but drop the explicit reference to 
their temperature-dependence.

Let us now add the effective interactions between the rectangular 
variants at different 
locations.
These are  also induced by elastic effects, via compatibility constraints \cite{Lookman01}, 
but can be modelled 
by an effective spin-interaction
\begin{equation}
H_I=-\sum_{(ij)} J({\bi{R}}_{ij}) S_{i}S_{j}
\label{eq:effective_interaction_Hamiltonian}
\end{equation}
where $(ij)$ indicates a pair of sites, ${\bi{R}}_{ij}$ is the 
separation of `sites` $(i,j)$ and $J({\bi{R}})$ has 
a short-range attractive part (arising from the energetic costs of 
strain gradients)  and also long-range 
contributions of $\pm$ sign depending upon the orientation of  ${\bi{R}}_{ij}$,
together causing frustration \cite{Lookman01} \footnote{Strictly we should also add short-range terms penalising the juxtaposition of $S_i =0$ and $S_i =\pm1$, for example of the form $K\sum_{\rm n.n.} S_{i}^2 (1 -S_{j}^2)$, and favouring neighbouring $S=0$, for example $-L\sum_{\rm n.n.} (1-S_{i}^2)(1-S_{j}^2)$.}.

Overall the 
behaviour can be considered as determined by the total effective 
`Hamiltonian', $H=H_0 +H_I$, together with appropriate
boundary and further thermal effects.

\subsubsection{Homogeneous case}

It is instructive to consider first 
the homogeneous case, with all $D$ equal. At high temperatures, 
where the $D$ are all positive, the  $H_0$ term alone would lead 
to all $S_{i}=0$, corresponding to austenite, whereas for all $D$ 
negative $H_0$ alone yields an Ising $S_{i}=\pm1$ behaviour 
with the $H_I$ term 
ordering the rectangular distorsions so as to minimise the 
overall energy. 
The very experimental 
existence of twins demonstrates a type of `anti-ferromagnetism'; its
specific form is a consequence of the combination of the
long-range decay and the 
anisotropic variation of sign of the interactions. 
More specifically the long-range 
interaction 
in two dimensions behaves dominantly as \cite{Lookman03}
\begin{equation}
J({\bi{R}}_{ij}) \propto - \cos{4\theta_{ij}}/{\mid {\bi{R}}_{ij} \mid}^2
\label{eq:effective_interaction}
\end{equation}
where $\theta_{ij}$ is the polar angle of ${\bi{R}}_{ij}$.  
The anisotropic variation is thus ferromagnetic 
along the ${\pi}/4$ and $3{\pi}/4$ directions and
antiferromagnetic in the $0$ and ${\pi}/2$ directions.
Minimization of the two energetic terms immediately suggests the 
formation of alternating twinned stripes of internally ferromagnetic 
`spins', the stripes all oriented along either $\pi/4$ or $3\pi/4$ but 
with the rectangular variants alternating from stripe to stripe between 
vertical and horizonal long axis. This order 
is so as to take advantage of the ferromagnetic interactions along the 
in-plane direction while minimising the cost of the antiferromagnetic 
interactions along the 
$x$ and $y$ directions. If we denote the energy contributions (i) 
along the direction of 
the twin stripes, $E_1$; (ii) perpendicular to these stripes, $E_2$; 
(iii) along $x$, $E_3$; (iv) along $y$, $E_4$, then we find $E_1$ and 
$E_2$ negative, with 
$\mid E_1 \mid >> \mid E_2 \mid \approx \mid E_{3} \mid = \mid E_{4} \mid $ 
\footnote{Along 
a stripe all the interactions are ferromagnetic and so add coherently; 
perpendicular to the stripe planes the alternating-variant planes 
interfere destructively and the long range of the interaction leads 
to a significant reduction of the binding energy in this direction as 
compared with in-plane; in the $0$  and $\pi/2$ directions there is a 
combination of both long-range antiferromagnetic interaction and 
plane-spin sign alternation; the short-range ferromagnetic 
interaction favours planes of more than single-spin width.}. 
The balances of gains and losses of energy and entropy, including  boundary effects 
determine optimal twin widths. 

In fact 
it is unnecessary for the $D$ to be negative for this 
twinned order to occur since the energy gained from $H_I$ by 
the ordering of the $S_{i}=\pm 1$ can (and typically does) 
outweigh the costs 
associated with 
$H_0$ even for positive  $D$, up to a critical value, in 
analogy with the phenomenon of induced moment 
behaviour in certain magnetic systems \cite{Trammel63}. The
transition from ground state 
$S_{i}=0$ 
to $S_{i}=\pm1$ would then be 
first order.  If the $D$-variation were independent of 
thermal effects and of the inter-site ordering, this qualitative 
feature would continue as the temperature 
is increased, until a tricritical point is reached. Note, however,
 that in 
the picture being painted here temperature plays two roles, one in effectively
tuning $D$ and another in the cooperative thermal ordering of the 
effective `spins'. There is no reason for the  temperature-scales,
or `transition temperatures', of the two 
effects to be simply related but experimental experience suggests 
that in practice a tricritical temperature is not 
normally reached while $D(T)$ is negative.

\subsubsection{Inhomogeneous case}

Allowing for inhomogeneity among the $D_{i}$ permits the 
possibility of an intermediate ordered phase between austenite 
and periodically twinned phase, as will now be shown.

For preliminary mental orientation, let us first consider a 
situation in which the $D_i$ take just two 
values randomly, one large and positive and the other large and negative.
The model Hamiltonian, 
\begin{equation}
H= \sum_{i} D_i(T) S_{i}^2 - \sum_{(i,j)} J({\bi{R}}_{ij}) S_{i}S_{j}:{\qquad} S_{i} = 0,\pm 1
\label{eq:effective_Hamiltonian}
\end{equation}
is then recognised as essentially analagous to that  
for a site-disordered Ising spin glass of magnetic and non-magnetic atoms,
\begin{equation}
H= -\sum_{(ij)} c_{i} c_{j} J({\bi{R}}_{ij}) S_{i}S_{j}: {\qquad} S_{i} =\pm 1,{\qquad} c_i =1,0
\label{eq:effective_sg_Hamiltonian}
\end{equation}
where the $c_i =1$ indicate magnetic sites, $c_i=0$  non-magnetic; ignoring bootstrap effects,
the magnetic sites 
correspond to those
with $D_i$ negative and the non-magnetic sites to $D_i$ positive. The detailed
make-up of attractive and repulsive forces is rather different from
those of either conventional metallic or conventional semiconducting spin glass 
systems \cite{Mydosh} but experience has taught that these details are not normally crucial to 
the existence of 
a spin glass phase. (The energy scale of the present pseudo-spin system is also different from that of real spin systems; much higher.) Based on 
this analogy one is led to conclude that 
the ground state will be antiferromagnetic/twin-stripe-ordered, `spin glass'/tweed or `paramagnetic'/austenite depending upon the
relative concentration of sites with the two signs of $D$; twin-ordered for 
a sufficient concentration of negative $D$, spin glass beneath this critical concentration and above a lower percolation-frustration threshold, beneath which no cooperative order is 
possible\footnote{Note that the existence of a finite lower threshold to paramagnetism 
depends on the character and range of the longer-ranged interactions. If ferromagnetic and extending to infinite-range there would be no zero-temperature threshold while with isotropic antiferromagnetic interactions it is believed that there is no long-range order in the limit of small concentration, even when those interactions are long-ranged \cite{Bhatt, McLenaghan}. Of course, at finite temperature (\ref{eq:effective_sg_Hamiltonian}) always has a lower concentration ordering threshold.}. 

A more realistic distribution of $D$ would be a continuous but bounded 
one; 
say of width $\Delta{D}$ around a mean of $D_0$, with $D_0$ effectively 
decreasing
monotonically (across zero) with 
reducing temperature $T$ of the underlying martensitic alloy.
At high enough $D_0$ the 
ground state would be non-magnetic ${S_{i}=0}$ (austenite). 
Clearly, any site whose $D$-value is negative would contribute an effective 
magnetic site to (\ref{eq:effective_sg_Hamiltonian}) and consequently
at low enough $D_0$ all sites would be `magnetic` and twin-ordered.
But for sufficient finite $\Delta{D}$  an intermediate spin 
glass(tweed) state may be expected.   

Let us first consider the case in which bootstrapping effects are ignored. 
Then the effective magnetic concentration would be equal to the number 
of sites with negative $D$ and cooperative order would onset at 
the corresponding lower critical concentration of a cooperative phase of  
(\ref{eq:effective_sg_Hamiltonian}).
For a continuous $D$-distribution 
this concentration would grow continuously as $D_0$ is reduced, so that the transition from non-magnetic (austenite) will
be to spin glass via a continuous transition. Eventually, as $D_0$ is reduced 
the concentration will reach that at which twin-order becomes 
preferable to spin glass and a further transition would occur. 

However, when the effects of bootstrapping of moments is included both these transitions could
become first order, the gain in interaction binding energy overcoming the cost 
of finite $S_i$ on certain sites even when their corresponding $D_i$ is still positive, up to a self-consistently determined $D_{c}$. This is 
possible for austenite to tweed whenever the low-concentration limit for tweed order of (\ref{eq:effective_sg_Hamiltonian}) is non-zero. The tweed-twin transition 
will be first order when at the critical concentration separating tweed and twin phases of (\ref{eq:effective_sg_Hamiltonian}) the slope of the transition temperature versus concentration of the
twin phase is sufficiently greater than that in the tweed phase that a jump in effective concentration (due to discontinuous moment formation) becomes energetically advantageous. Of course if the statistical inhomogeneity, 
and therefore $\Delta D$, is too small the tweed phase could be bypassed in the 
discontinuous `moment' jump.

\subsection{Symmetry-breaking fields}

Above we have not discussed explicitly the inclusion of effects breaking the symmetry between alternative variants. These can occur due to either external  or internal stresses and would add terms of the form
\begin{equation}
H_{Stress}= -\sum_{i} h_{i} S_{i}
\label{eq:symmetry-breaking_Hamiltonian}
\end{equation}
with contributions to the $h_{i}$ either externally imposed or a consequence of internal inhomogeneities favouring particular local orientations. Thus one could 
extend the Hamiltonian to include these terms and model with
\begin{equation}
h_{i}= h_{i}^{imposed} + h_{i}^{random}
\label{eq:symmetry-breaking_fields}
\end{equation}
with the $h_{i}^{random}$ randomly chosen from some distribution $P(h^{random})$.

\subsection{Three dimensions}
Let us now turn to three dimensions, the more normal dimensionality of martensitic systems. Again first order transitions from high to lower local symmetry are 
observed experimentally; in classic cubic systems with the two types 
of rectangular lower-symmetry local variants of the two dimensional example replaced
by three tetragonal variants, orientated orthogonally along the $x,y$ and $z$ axes.
Again the interaction terms lead to cooperative spatial ordering of the variants. At low temperatures this ordering is again of complementary twins, implying again 
long-range interactions of oscillating sign as a function of angular orientation \cite{Lookman01}. Both of these features can be emulated by extensions of the simple spin model used above or by an axis-director analogue of it. For example the three tetragonal variants can be described analagously to liquid-crystal (axis-)`directors', with interactions between them analogous to those in quadrupolar glasses \cite{Goldbart}. Local inhomogeneities again lead to different local propensities to cubic (zero director) or tetragonal (finite director)
and hence to
random variations in the effective concentration of directors. This in turn can be anticipated to permit the insertion of a quasi-random three-dimensional tweed phase between austenite and twinned phases. A possible model formulation would be to employ the 
effective Hamiltonian 
\begin{equation}
H=\sum_{i} D_i(T) {\mid \bi{S_{i}}\mid}^2 -\sum_{(i,j)} {\mid {\bi{S}}_i \mid} {\mid {\bi{S}}_j \mid}   
J({\bi{R}}_{ij}) 
(2({\bi{S}}_{i} . {\bi{S}}_{j})^2 -1)
\label{eq:3d_Hamiltonian}
\end{equation}
where the spins are restricted to
${\bi{S_{i}}} = 0, {\bf{\hat{x}}},{\bf{\hat{y}}}$,or ${\bf{\hat{z}}}$ and the 
last three options are respectively unit vectors in the $x$, $y$ and $z$ directions. 
Another possible description would be in terms of a four-state Potts spin 
glass \cite{ES} in a field and with anisotropic exchange; one of 
the Potts dimensions corresponding to austenite, the other three to the 
orthogonal variants, with the exchange interactions only among these 
latter three dimensions. 

\section{Properties}
Among the most interesting properties of martensitic systems is that of shape-memory. In so-called `one-way shape-memory' a system which is forged into a particular shape at a high temperature, in the austenite phase, cooled and then distorted, will regain the original shape on re-heating. This can be explained easily from twinning alone along with the feature of ease of distorsion in the twinned phase \cite{Bhattacharya}. The initial distorsion corresponds to the imposition of a boundary (at whatever necessary energy cost). Cooling alone retains that shape, with the twin stripe-widths accommodating it. In the twinned phase, however, the shape is easily modified at low energy cost by adjusting the individual twin section widths by moving the intertwin boundaries. Reheating removes the twins and their boundaries and the sample regains the originally imposed shape. There is however often observed another type of shape memory, `two-way-shape memory' in which the system remembers its history both on heating and on cooling \cite{Experiment}. This is not readily explainable by twinning alone. However,  such two-way memory is a characteristic feature of spin glassses \cite{Young, Vincent} and hence is attributable to tweed \cite{Sethna}. The present model suggests a way to model it minimally, but further pursuit of its study is deferred to a future paper.

Another of the characteristic features of a spin glass is preparation-dependence, as illustrated in a difference between the susceptibility as measured by first cooling from the paramagnetic state and then applying a field (ZFC: zero-field cooling) and that observed
by first applying the field and then cooling (FC: field-cooling), demonstrating non-ergodic behaviour over experimental times. There should be a corresponding difference in uniaxial compliances (or, inversely, compressibilities) in tweed measured along principal axes\footnote{ Note that uniaxial compression is needed so that the effective field 
has a different energetic 
consequence for the different types of tetragonal distorsion;
isotropic compressibility
would not have the same effect since it is 
an ``irrelevant field'' for separating tetragonal twins. Neither would compressibility at $\pi/4$ to the tetragonal variant long directions.}. This effect is a consequence of the multiplicity of metastable states of spin glasses, their non-trivial evolution under changes of control parameters (such as temperature or applied field) and consequental slow dynamics in exploring all of phase space \cite{Mezard,Nishimori}. It should however also be noted that the normal fluctation-dissipation relation does not hold for spin glasses \cite{Young} and a simple measurement of the sound velocity will correspond more to the ZFC situation. However, variations can be expected as a function of frequency.

We might also note in passing that the full Gibbs susceptibility of a mean-field spin glass (and, at least to a good approximation, the FC susceptibility of a real spin glass) is independent of temperature and hence one might anticipate a similar independence in a martensitic alloy throughout the tweed regime. 
 
Aging behaviour in glasses has a long history \cite{Struik} and similar aging features should exist in tweed. More recent work on spin glasses has shown how many aging quantities can be expressed as a function of the ratio of times $t/t_w$ where $t_w$ is the time following a rapid quench and $t$ is time of measurement, for both $t$ and $t_w$ large. A corresponding systematic exploration of tweed would seem to be called for.

Another 
intriguing characteristic of conventional spin glasses is rejuvenation 
\cite{Vincent},
whereby  a system  relaxing/aging in an external 
d.c. field  appears 
to start
relaxing/aging anew when that field is suddenly changed, ignoring its previous aging; this 
feature is, for example, apparent in an observation of the a.c. 
out-of-phase susceptibility. Again one would expect an analogue for tweed. 
 
\subsection{Relation to prior models}

The models considered above can be viewed as extensions of the 
Blume-Capel \cite{Blume} or Blume-Emery-Griffiths models \cite{BEG}; in two dimensions to allow for randomness of the local anisotropy term and long-range and orientational sign variation of the exchange interaction with separation 
${\bf{R}}_{ij}$, and in three dimensions also to allow for Potts-like director replacement of the BEG $S=1$ Ising spins. A complementary discrete random field Potts modelling of elastic systems has recently been proposed by Cerruti and Vives \cite{Cerruti}.

\section{Soluble models}

Although the physical conclusions indicated above seem inevitable, 
the actual models introduced are almost certainly not completely soluble, 
as neither is
the corresponding equation ({\ref{eq:effective_sg_Hamiltonian}}) of an even-simplified conventional site-disordered spin glass. 
In theoretical and simulational studies of conventional spin glasses
one usually replaces site disorder with bond disorder 
$P(J_{ij})$, as suggested by Edwards and Anderson in their classic seminal 
paper \cite{EA}.  Taking the mean $J_{ij}$ to be non-zero 
\cite{Sherrington-Southern} 
permits competition (and transition) between two types of frozen phase, 
spin glass and ferromagnet (or anti-ferromagnet). An exactly soluble version is
the infinite-range extension of Sherrington and Kirkpatrick 
\footnote{SK can be extended to cover antiferromagnetic ordering via the 
introduction of two sublattices as in \cite{KS}.}\cite{SK} \cite{Sethna92}.
A tempting analogue in the present case is a 
variant of the Ghatak-Sherrington model \cite{GS} (see also 
\cite{Sellitto} \cite{CL}), with 
\begin{equation}
H_{GS}=\sum_{i} D_i S_{i}^2 + \sum_{(i,j)} J_{ij} S_{i}S_{j}: S_{i} = 0,\pm 1
\label{eq:GS2_Hamiltonian}
\end{equation}
where the $J_{ij}$ are independently distributed according 
to some $P(J_{ij})$, including allowance for non-zero mean,
and the $D_i$ also independently according to some other 
distribution $P_{D}(D)$. Such a model should be straightforwardly soluble by the 
methods developed for the SK model and its extensions \cite{Mezard,Nishimori} but 
it would miss many of the crucial features of the martensitic systems (including the feature of easily malleable twin planes). Consequently computer simulation of the model system seems to be the sensible next direction to pursue at the theoretical level.

\section*{Conclusions}

A simple spin-glass-like modelling has been presented emulating key ingredients of martensitic alloy transformations. Qualitative considerations and analogies have been
used to suggest that tweed should be viewed as a spin glass, echoing a message originally expressed, via a different formulation, by Kartha et. al.\cite{Kartha91}. 
Analogies with aging, rejuvenation and memory of spin glasses suggest explanations and experimental investigations of the behaviour of martensitic alloys. The study also exposes a new type of spin glass Hamiltonian worthy of further investigation. 

This brief paper has concentrated on devising simple models and considering their likely thermodynamic phases and transitions, together with some anticipated properties by analogy with more conventional spin glasses. These need further more rigorous consideration,  but so also does the dynamics of these systems viewed within the simple spin-Hamiltonian modelling. There has already been some study of the dynamics within a fuller elastic strain formulation without alloy-inhomogeneity \cite{Lookman03} ({\it i.e.} without imposed disorder quenched into the controlling equations) and for a different simple model with quenched disorder \cite{Kartha95}. A start has also been made in terms of computer dynamical iteration of the mean field solution for the two-dimensional homogeneous-Hamiltonian case \cite{Lookman08}. However, a fuller examination along the lines of those employed for more conventional spin glasses is now called for, with special attention to the metastabilities that characterise systems with first order transitions; note that conventional spin glasses  have thermodynamically continuous transitions even though soluble model systems with $p>2$-spin interactions or with interactions lacking symmetry of definiteness (such as Potts or quadruplor) spins have discontinuous replica symmetry breaking (DRSB) spin glass onset. We might also note that structural glasses, which have no imposed quenched localised disorder in their Hamiltonians, have quasi-transition temperatures  $T_g$ (glass transition - high viscosity) and Kauzman temperatures $T_K$ (temperature at which the entropy difference between liquid and glass extrapolates to zero) analogous to the dynamical and thermodynamic glass transitions of a DRSB model. Hence it is also possible that an analagous dynamically-self-generated quasi-disorder could occur in martensites \cite{Lookman08}. Simple modelling such as introduced here might provide a useful laboratory within which to probe the physics.

\section*{Acknowledgements}

This work was stimulated by discussions over a few summers with Turab Lookman and Avadh Saxena at Los Alamos National Laboratory and I would like to thank them both, also for recent advice during the writing of this paper. I thank James Sethna for drawing my attention to refs.\cite{Sethna}\cite{Sethna92} and also for helpful comments on a draft of this paper. Financial support was provided by EPSRC through grant GR/R83712/01 and by the US Department of Energy through CNLS, Los Alamos National Laboratory. The Santa Fe Institute also provided hospitality during the germination and the writing of this paper.

\section*{References}

%
\end{document}